\documentclass[aps,pre,twocolumn,superscriptaddress]{revtex4}

\pdfoutput=1

\usepackage{graphicx}

\begin{document}


\title{Effect of spatial structure on the evolution of cooperation}


\author{Carlos P.\ Roca}
\email[]{cproca@math.uc3m.es}
\homepage[]{http://www.gisc.es}
\affiliation{GISC/Departamento de Matem\'aticas, Universidad Carlos III de
Madrid, Spain}

\author{Jos\'e A.\ Cuesta}
\affiliation{GISC/Departamento de Matem\'aticas, Universidad Carlos III de
Madrid, Spain}

\author{Angel S\'anchez}
\affiliation{GISC/Departamento de Matem\'aticas, Universidad Carlos III de
Madrid, Spain}
\affiliation{Instituto de Biocomputaci\'on y F\'isica de Sistemas Complejos
(BIFI), Universidad de Zaragoza, Spain}
\affiliation{Instituto de Ciencias Matem\'aticas CSIC--UAM--UC3M--UCM,
Madrid, Spain}



\begin{abstract}
Spatial structure is known to have an impact on the evolution of cooperation,
and so it has been intensively studied during recent years. Previous work has
shown the relevance of some features, such as the synchronicity of the updating,
the clustering of the network or the influence of the update rule. This has been
done, however, for concrete settings with particular games, networks and update
rules, with the consequence that some contradictions have arisen and a general
understanding of these topics is missing in the broader context of the space of
$2 \times 2$ games. To address this issue, we have performed a systematic and
exhaustive simulation in the different degrees of freedom of the problem. In
some cases, we generalize previous knowledge to the broader context of our study
and explain the apparent contradictions. In other cases, however, our
conclusions refute what seems to be established opinions in the field, as for
example the robustness of the effect of spatial structure against changes in the
update rule, or offer new insights into the subject, e.g.\ the relation between
the intensity of selection and the asymmetry between the effects on games with
mixed equilibria.
\end{abstract}



\maketitle


\section{Introduction}

\label{sec:intro}

Cooperation is a key force in evolution, present in all
scales of organization, from unicellular organisms to complex modern human
societies \cite{maynard-smith:1995}. For this reason the elucidation of the
emergence and stabilization of cooperative behavior has become a core problem in
biology, economics and sociology \cite{pennisi:2005}. Evolutionary game theory
has proven to be one of the most fruitful approaches to investigate this
problem, using evolutionary models based on so-called social dilemmas
\cite{maynard-smith:1982,axelrod:1984}. Several mechanisms have
been proposed to explain the appearance and survival of cooperation
\cite{nowak:2006b}, the structure of the population being one of them, which in
this context is also referred to as ``network reciprocity''. The presence of
structure means that each individual does not interact with every other, but
with a small subset of the population, which constitutes her neighborhood and
is arranged according to an underlying network of relationships. This idea was
very successfully introduced by Nowak and May in their seminal
paper \cite{nowak:1992}, stimulating a wealth of work that continues to date
(see \cite{szabo:2007} for a review).

The current view on the influence of spatial structure, as a particular case of
population structure, is that in general it promotes cooperation, given the
positive effects reported on \cite{nowak:1992} on the most demanding $2 \times
2$ game, namely the Prisoner's Dilemma \cite{axelrod:1981}. Many other models
have confirmed this beneficial effect \cite{szabo:2007}, with the only exception
of anti-coordination games such as Hawk-Dove or Snowdrift
games \cite{hauert:2004}.
Most studies, however, have concentrated on the Prisoner's Dilemma
\cite{nowak:1992,nowak:1994,lindgren:1994,hutson:1995,grim:1996,nakamaru:1997,
szabo:1998,brauchli:1999,abramson:2001,cohen:2001,vainstein:2001,lim:2002,
schweitzer:2002,ifti:2004,tang:2006,perc:2008}, and the effect of spatial
structure on other games has received much less attention, as is shown by the
much smaller number of studies on Snowdrift
\cite{killingback:1996,hauert:2004,sysi-aho:2005,kun:2006,tomassini:2006,
zhong:2006} and specially on Stag Hunt
games \cite{blume:1993,ellison:1993,kirchkamp:2000} (these lists of references
are by no means exhaustive; we refer the interested reader to the review
\cite{szabo:2007} and also to the supplementary material of \cite{wang:2008}).
The importance of considering other kind of games besides Prisoner's Dilemma
should not be underestimated, as they may be essential in biological
or economic applications. Moreover, it has been recently shown 
\cite{fort:2008} that, if the game itself is subject to evolutionary forces,
Stag Hunt games may be of special relevance. Besides this, previous research has made use of a variety of rules for the update of strategies, combined with
different network topologies, with the result that it is currently impossible to discern for most cases whether the reported effects on the evolutionary outcome are caused by the population structure, the update rule or the combination of
both.

To provide a comprehensive picture of this issue, we have performed
a thorough and systematic computational study, taking into account
the most important symmetric $2 \times 2$ games and update rules on a number of
degree-homogeneous network models, including two-dimensional regular
lattices, which are by far the most prototypical models for spatially extended
populations. The reader should note that there have been previous attempts at
achieving the same goal, the most prominent among which is a work by Hauert
\cite{hauert:2002}.
Despite the virtues of this paper, we consider that two important
shortcomings render it a preliminary and inconclusive attempt: (1) The
results reported in \cite{hauert:2002} for stochastic update rules
are not accurate, because the time of convergence is more than one order of
magnitude smaller than the minimum required for the system, given its size,
to reach a stationary state (see Appendix); and (2) the
effect of spatial structure is evaluated comparing with a well-mixed population,
which makes it impossible to discern if the influence under study is due to the
spatial distribution of neighbors or just to the mere \emph{context preservation}
\cite{cohen:2001} of a degree-homogeneous random network (this distinction is
crucial, as it reveals the role of the clustering coefficient in spatial
networks that we will show in Sec.~\ref{sec:results}). In contrast, to
assess the influence of a given spatial structure we have compared the results
not only with the unstructured (i.e.\ well-mixed) population, but also with the
homogeneous random population of the same degree, and we have introduced a
quantitative measure for the cooperation achieved on each kind of game. In
addition, we have considered in detail the time evolution, both under
synchronous and asynchronous update schemes, in order to understand the
fundamental dynamical mechanisms involved, and we have addressed the issue of
the influence of selection pressure.

As a result of this exhaustive study we have reached a number of conclusions
that must be put in the context of previous research. In some cases, these
are generalizations of known results to wider sets of games and update rules, as
for example for the issue of the synchrony of the updating of strategies
\cite{nowak:1992,huberman:1993,nowak:1994,lindgren:1994,kirchkamp:2000,
kun:2006,tomassini:2006} or the effect of small-world networks vs regular
lattices
\cite{abramson:2001,masuda:2003,tomochi:2004,tomassini:2006}. In
other cases, the more general view of our analysis allows to integrate
apparently contradictory results in the literature, as the cooperation on
Prisoner's Dilemma vs.\ Snowdrift games
\cite{nowak:1992,killingback:1996,hauert:2004,sysi-aho:2005,tomassini:2006}, or the importance of clustering in spatial lattices
\cite{cohen:2001,ifti:2004,tomassini:2006}. Other conclusions of ours, however,
refute what seems to be established opinions in the field, as the alleged
robustness of the positive influence of spatial structure on Prisoner's Dilemma
\cite{nowak:1992,nowak:1994,hauert:2002}. And finally, we have
reached additional conclusions, such as the robustness of the influence of spatial
structure on coordination games, and the asymmetry between the effects on
games with mixed equilibria (coordination and anti-coordination games) and how
it varies with the intensity of selection.

In Sec.~\ref{sec:evol-games} we give the background in evolutionary
game theory relevant to our work, introducing symmetric $2 \times 2$ games and
several evolutionary rules for the update of strategies.
Section~\ref{sec:previous-results} recalls some results in the
literature, highlighting the unresolved questions that motivated this
study. In Sec.~\ref{sec:results} we present our results about the influence of
spatial structure on the evolution of cooperation, addressing these questions
and detailing our conclusions on the topic. Subsequently,
Sec.~\ref{sec:discussion} deals with the basic mechanisms that underlie the
influence of spatial structure. Finally, Sec.~\ref{sec:conclusions}
concludes the paper summarizing our most important points and giving some
additional remarks.
One Appendix is included, which describes the technicalities of the
computer simulations.

\section{Evolutionary games}

\label{sec:evol-games}

Let us consider a symmetric $2 \times 2$ game, a game with two players who choose
between two strategies and with no difference in role. Each player obtains a
payoff given by the following matrix
\begin{equation}
\label{eq:payoff-matrix}
\begin{array}{cc}
  & \begin{array} {cc} \mbox{C} & \mbox{D} \end{array} \\
  \begin{array}{c} \mbox{C} \\ \mbox{D} \end{array} &
  \left( \begin{array}{cc} 1 & S \\ T & 0 \end{array} \right).
\end{array}
\end{equation}
The rows represent the strategy of the player who obtains the payoff and the
columns that of her opponent. We will come back to this choice of parameters at
the end of this Section, after introducing the evolutionary rules.

The strategies are labeled C and D for cooperate and defect,
because we interpret the game as a social dilemma. Indeed, certain
values of $S$ and $T$ undermine a hypothetical situation of mutual
cooperation. If $S<0$ a cooperator faces the risk of losing if the
other player defects, performing worse than with mutual defection. If
$T>1$ a cooperator has the temptation to defect and obtain a payoff
larger than that of mutual cooperation. Both tensions determine the
social dilemmas represented by symmetric $2 \times 2$ games
\cite{macy:2002}. Restricting the values of the coefficients within
the intervals $-1 < S < 1$ and $0 < T < 2$, we have the Harmony game
\cite{licht:1999} ($0 < S,\, T < 1$) and three classic social
dilemmas: the Prisoner's Dilemma \cite{axelrod:1981} ($-1 < S <
0,\, 1 < T < 2$), the Stag Hunt game \cite{skyrms:2003} ($-1 < S
< 0 < T < 1$), and the Hawk-Dove \cite{maynard-smith:1973} or
Snowdrift game \cite{sugden:2004} ($0 < S < 1 < T < 2$). Each
game corresponds, thus, to a unit square in the $ST$-plane.

There is experimental evidence of interactions that basically correspond to the above stylized games. In general, Prisoner's Dilemmas or Snowdrift games are more frequently found in the dominance or co-existence among different traits or species in biological contexts \cite{turner:1999,gore:2009}, whereas Stag Hunt games are more related to the problem of coordination or equilibrium selection from an economical viewpoint \cite{harsanyi:1988}. We refer the interested reader to the list of references of \cite{gore:2009} for the former case and to \cite{camerer:2003} for the latter.

\begin{figure*}[t!]
\centering
\includegraphics[width=0.99\textwidth]{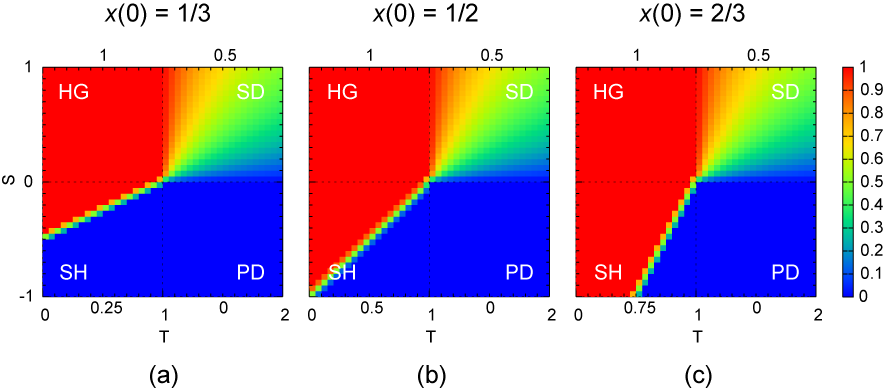}
\caption{(Color online) Asymptotic density of cooperators $x^*$ in a complete network with
the replicator rule as update rule, starting with different initial densities of cooperators $x^0$, of values 1/3 (a), 1/5 (b) and 2/3 (c). Each unit square corresponds to a game: Harmony upper-left, Prisoner's
Dilemma lower-right, Stag Hunt lower-left, Snowdrift upper-right. The outcome is the same as that of the standard replicator dynamics on a well-mixed population (see text). Notice that Stag Hunt games are the only ones whose outcome depends on the initial condition, because of their bistable character. The values of the global measure of cooperation, obtained as the mean value over each game region, are shown next to each game square (above in the case of Harmony and Snowdrift games, and below for Stag Hunt games and Prisoner's Dilemmas).}
\label{fig:compnet}
\end{figure*}

To study the competition between cooperation and defection from an
evolutionary perspective, the payoffs obtained by playing the game
are considered as fitness \cite{maynard-smith:1982} and a Darwinian dynamics is
introduced to promote the fittest strategy. The classic framework to do so is
the replicator dynamics \cite{hofbauer:1998,gintis:2000}, which assumes
an infinite and well-mixed population, i.e.\ a population with no
structure, where each individual plays with every other. Let $x$ be
the density of cooperators, and $f_c$ and $f_d$ the fitness of a
cooperator and a defector, respectively. The replicator dynamics
posits that $x$ evolves as \cite{hofbauer:1998}
\begin{equation}
\label{eq:repldyn-wellmixed}
\dot{x} = x (1-x) ( f_c - f_d ) .
\end{equation}
Then, if cooperators are doing better than defectors their density rises
accordingly, and the opposite occurs if they are doing worse. Provided that the
initial density of cooperators $x^0$ is different from 0 and 1, the asymptotic
state of this dynamical system is, for each game ($x^*$ represents the
asymptotic density of cooperators) \cite{hofbauer:1998}: Harmony,
full cooperation, $x^* = 1$; Prisoner's Dilemma, full defection, $x^* = 0$; Stag
Hunt, full cooperation if $x^0 > x_e$, or full defection if $x^0 < x_e$;
Snowdrift, mixed population with $x^* = x_e$, regardless of the initial density
$x^0$. For both Stag Hunt and Snowdrift games the mixed equilibrium
$x_e$ has a value
\begin{equation}
\label{eq:mixedeq}
x_e = \frac{S}{S+T-1} .
\end{equation}
Notice that this mixed equilibrium is unstable in Stag Hunt games whereas it is (globally) stable in Snowdrift games. It is important to note also that the outcomes of the four games presented encompass all the possible equilibrium structures of any symmetric $2 \times 2$ game
\cite{rapoport:1966,roca:2006}.

The standard equivalent version of this evolutionary model, for finite
populations and discrete time, is built by placing the population on a
complete network and by making use of the following
rule for the update of strategies, known as the \emph{replicator
rule} or \emph{proportional imitation rule} \cite{helbing:1992,schlag:1998}. Let
$i = 1 \ldots N$ label the individuals in the population. Let $s_i$ be the
strategy of player $i$, $\pi_i$ her payoff and $N_i$ her neighborhood, with
$k_i$ neighbors. With the replicator rule one neighbor $j \in N_i$ is chosen
at random. The probability of player $i$ adopting the strategy of player $j$
at time $t$ is given by
\begin{equation}
\label{eq:replrule}
p^t_{ij} \equiv \mathcal{P}\{ s_j^t \to s_i^{t+1}  \} =
\left\{ \begin{array}{ll}
  ( \pi_j^t - \pi_i^t ) / \Phi \,, & \pi_j^t > \pi_i^t \,, \\
  0 \,, & \pi_j^t \leq \pi_i^t \,,
\end{array} \right.
\end{equation}
with $\Phi = \max(k_i,k_j) [ \max(1,T) - \min(0,S) ]$ to ensure that
$p^t_{ij} \in [0,1]$.

Figure~\ref{fig:compnet} shows the simulation results for this setting played
on a complete network, for different initial conditions. As expected, the results are in complete agreement with
the evolutionary outcome predicted by Eq.~(\ref{eq:repldyn-wellmixed}) for an
infinite well-mixed population. This outcome constitutes the reference against
which the effect on cooperation of a given population structure
will be assessed. Additionally, we introduce a quantitative measure
$\mathcal{C}_G$ for
the overall asymptotic cooperation in game $G$, given by the
average of $x^*$ over the corresponding region in the $ST$-plane. This global
index of cooperation has a range $\mathcal{C}_G \in [0,1]$ and
appears on the graphs by the unit square of each game.

Besides the replicator update rule, we have considered other imitative rules
that have received attention in previous research
\cite{szabo:2007}: the multiple replicator, Moran, unconditional
imitation and Fermi rules.

The \emph{multiple replicator rule} is similar to the replicator
rule, with the difference of checking simultaneously all the
neighborhood and thus making a strategy change more probable. According to
this rule, the probability that player $i$ maintains her strategy is
\begin{equation}
\label{eq:multireplrule}
\mathcal{P}\{ s_i^t \to s_i^{t+1} \} =
  \prod \limits_{j \in N_i} ( 1 - p_{ij}^t ) ,
\end{equation}
with $p^t_{ij}$ given by Eq.~(\ref{eq:replrule}). In case the strategy
update takes place, the neighbor $j$ whose strategy is adopted by
player $i$ is selected with probability proportional to $p^t_{ij}$.

With the \emph{Moran rule}, inspired on the Moran dynamics \cite{moran:1962}, a
player chooses the strategy of one of her neighbors, or herself's, with a
probability proportional to the payoffs
\begin{equation}
\label{eq:moranrule}
\mathcal{P} \{ s_j^t \to s_i^{t+1} \} =
  \frac {\pi_j^t - \Psi} {\sum \limits_{k \in N^*_i} ( \pi_k^t - \Psi )} ,
\end{equation}
with $N^*_i = N_i \cup \{i\}$.
Because payoffs may be negative in Prisoner's Dilemma and Stag Hunt games, the
constant $\Psi = \max_{j \in N^*_i}(k_j) \min(0,S)$ is subtracted from them.
Note that with this rule a player can adopt, with low probability, the strategy
of a neighbor that has done worse than herself.

Another frequently used rule is the \emph{unconditional imitation rule}, which
makes each player choose the strategy of the neighbor with the largest payoff,
provided this payoff is greater than the player's. This is a deterministic rule,
in contrast to all the others considered in this paper, which are stochastic.

Finally, an update rule that allows to investigate the influence of the
intensity of selection is the \emph{Fermi rule}
\cite{szabo:1998,traulsen:2006a}, based on the Fermi distribution function. With
this rule, a neighbor $j$ of player $i$ is selected at random (as with the
replicator rule) and the probability of player $i$ acquiring the strategy of
player $j$ is given by
\begin{equation}
\label{eq:fermirule}
\mathcal{P} \{ s_j^t \to s_i^{t+1} \} =
\displaystyle \frac {1} {1 + \exp \left( - \beta \, ( \pi_j^t - \pi_i^t )
\right)}.
\end{equation}
The parameter $\beta$ controls the intensity of selection, and can be understood
as the inverse of temperature or noise. Thus, low $\beta$ represents high
temperature or noise and, correspondingly, weak selection pressure. Again,
strategies performing worse can be chosen with this update rule.

Having introduced the evolutionary rules we will consider, it is important to
recall our choice for the payoff matrix (\ref{eq:payoff-matrix}), and discuss
its generality. The replicator rule and the unconditional imitation rule, in
degree-homogeneous networks like the ones considered in this work, are both invariant under translation and (positive) scaling of the payoff matrix, so for these rules (\ref{eq:payoff-matrix}) is the most general choice. Among the other
rules, the dynamics is also preserved for the multiple replicator rule, but it
changes upon translation for the Moran rule, or scaling for the Fermi rule. The
corresponding changes in these last two cases amount to a modification of the
intensity of selection, which we also consider in this work. Therefore we
believe that Eq.~(\ref{eq:payoff-matrix}) is general enough for our purposes.

From the viewpoint of statistical physics, these evolutionary models are
non-equilibrium systems that do not have a Hamiltonian and whose dynamics is
determined by local rules. Some studies have shown, for some particular cases,
that they belong to the Directed Percolation universality
class \cite{szabo:1998,szabo:2000}.

\section{Open questions in previous research}

\label{sec:previous-results}

First of all, we want to make clear that what follows does not
intend to be an exhaustive account of previous results in the literature.
We refer the interested reader to \cite{szabo:2007} for such a
detailed review. Our aim in this section is to present what are, from our point
of view, important results in the field, focusing only on those
that deal with the simplest cases of games on networks, namely symmetric $2
\times 2$ games on homogeneous fixed networks. These results form the basis
upon which one can address more sophisticated settings, such as those dealing 
with complex networks \cite{santos:2006a,gomezgardenes:2007}, co-evolution of
strategies and networks \cite{zimmermann:2004,pacheco:2006} or meta-evolution of
the update rules \cite{moyano:2009}, to name just a few of the options that are
being pursued by current research. While reviewing the available
literature we found difficulties in drawing general conclusions, as the
differences
between the models, regarding the game, the network or the update rule employed,
prevented a meaningful comparison of results. Moreover, some pending issues and
contradictions arose when we tried to integrate the main conclusions of
published work, motivating in the end the present study.

\begin{figure}
\centering
\includegraphics[width=7.5cm]{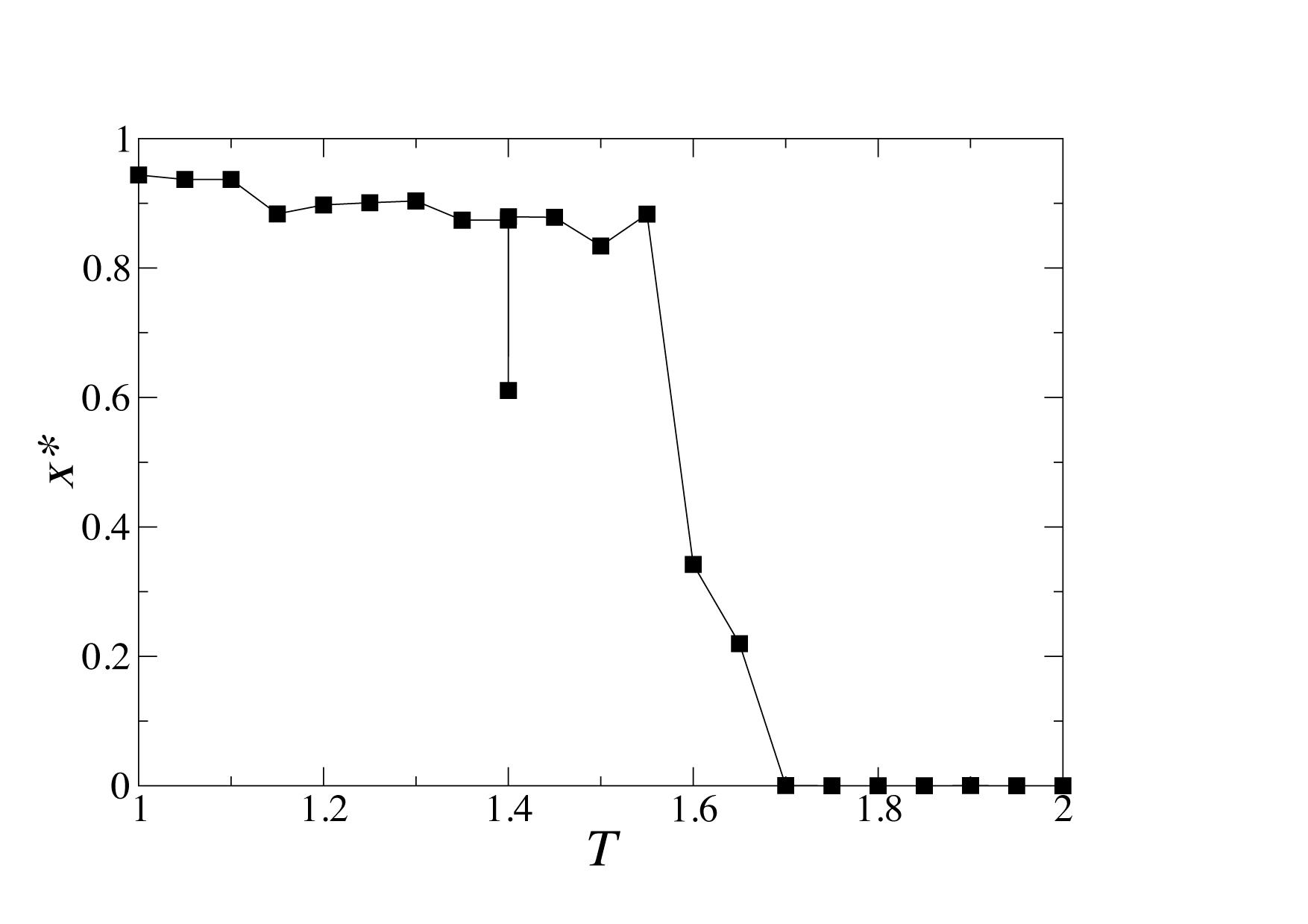}
\caption{Asymptotic density of cooperators $x^*$ in a square lattice with degree
$k=8$, when the game is the Prisoner's Dilemma defined by Eq.~(\ref{eq:nowak-pd})
($\epsilon=0$), as studied by Nowak and May \cite{nowak:1992} (update rule: unconditional imitation). Note that the
outcome with replicator dynamics on a well mixed population is $x^*=0$ for all
the displayed range of the temptation parameter $T$. Notice also the
singularity at $T=1.4$, with surrounding points located at $T=1.3999$
and $T=1.4001$ (see Fig.~\ref{fig:lattice-uncimit} and related discussion in Sec.~\ref{sec:results}). Lines are a guide to the eye.}
\label{fig:nowak}
\end{figure}

We begin with the pioneering paper published by Nowak and May in 1992
\cite{nowak:1992}, which showed the significant fostering of cooperation that
the spatial distribution of a population is able to produce in Prisoner's
Dilemma. This has become the prototypical example of the promotion of
cooperation favored by the structure of a population, also known as network
reciprocity \cite{nowak:2006b}. They considered the following Prisoner's Dilemma
\begin{equation}
\label{eq:nowak-pd}
\begin{array}{cc}
  & \begin{array} {cc} \mbox{C} & \mbox{D} \end{array} \\
  \begin{array}{c} \mbox{C} \\ \mbox{D} \end{array} &
  \left( \begin{array}{cc} 1 & 0 \\ T & \epsilon \end{array} \right),
\end{array}
\end{equation}
with $1 \le T \le 2$ and $\epsilon \lesssim 0$. Note that this one-dimensional
parametrization corresponds in the $ST$-plane to a region near the boundary
with Snowdrift games. Figure~\ref{fig:nowak} shows the great enhancement of
cooperation reported by \cite{nowak:1992}. The authors explained this influence
in terms of the formation of clusters of cooperators, which give cooperators
enough payoff to survive even when surrounded by (some) defectors.

A year later, Huberman and Glance \cite{huberman:1993} questioned the generality
of the results reported by Nowak and May \cite{nowak:1992}, in terms of the
synchronicity of the update of strategies, which is a very relevant issue in
biological contexts. Nowak and May used synchronous update, which means that
every player is updated at the same time, so the population evolves in
successive generations. Huberman and Glance, on the contrary, employed
asynchronous update (also called random sequential update),
in which individuals are updated independently one by one, so the neighborhood
of each player remains the same while her strategy is updated. They showed
that, for a particular game among those studied in \cite{nowak:1992} ---i.e.\
for a particular value of $T$ in Eq.~(\ref{eq:nowak-pd})---, the asymptotic
cooperation obtained with synchronous update disappeared. Nowak and May,
together with Bonhoeffer, published in 1994 a reply \cite{nowak:1994} to this
criticism,
defending the generality and robustness of the beneficial effect of spatial
structure on a variety of scenarios. Subsequent works have reinforced this
later viewpoint, restricting the effect reported by \cite{huberman:1993} to
particular instances of Prisoner's Dilemma \cite{lindgren:1994} or
to the short memory of players \cite{kirchkamp:2000}. Other works, however, in
the different context of snowdrift games \cite{tomassini:2006,kun:2006}
have found that the influence on cooperation can be positive or negative, in the
asynchronous case compared with the synchronous one. Then, an important
open question is: To which extent the synchronicity (or the lack of it) has an
influence, in general, on evolutionary games?

\begin{figure}
\centering
\includegraphics[width=7.5cm]{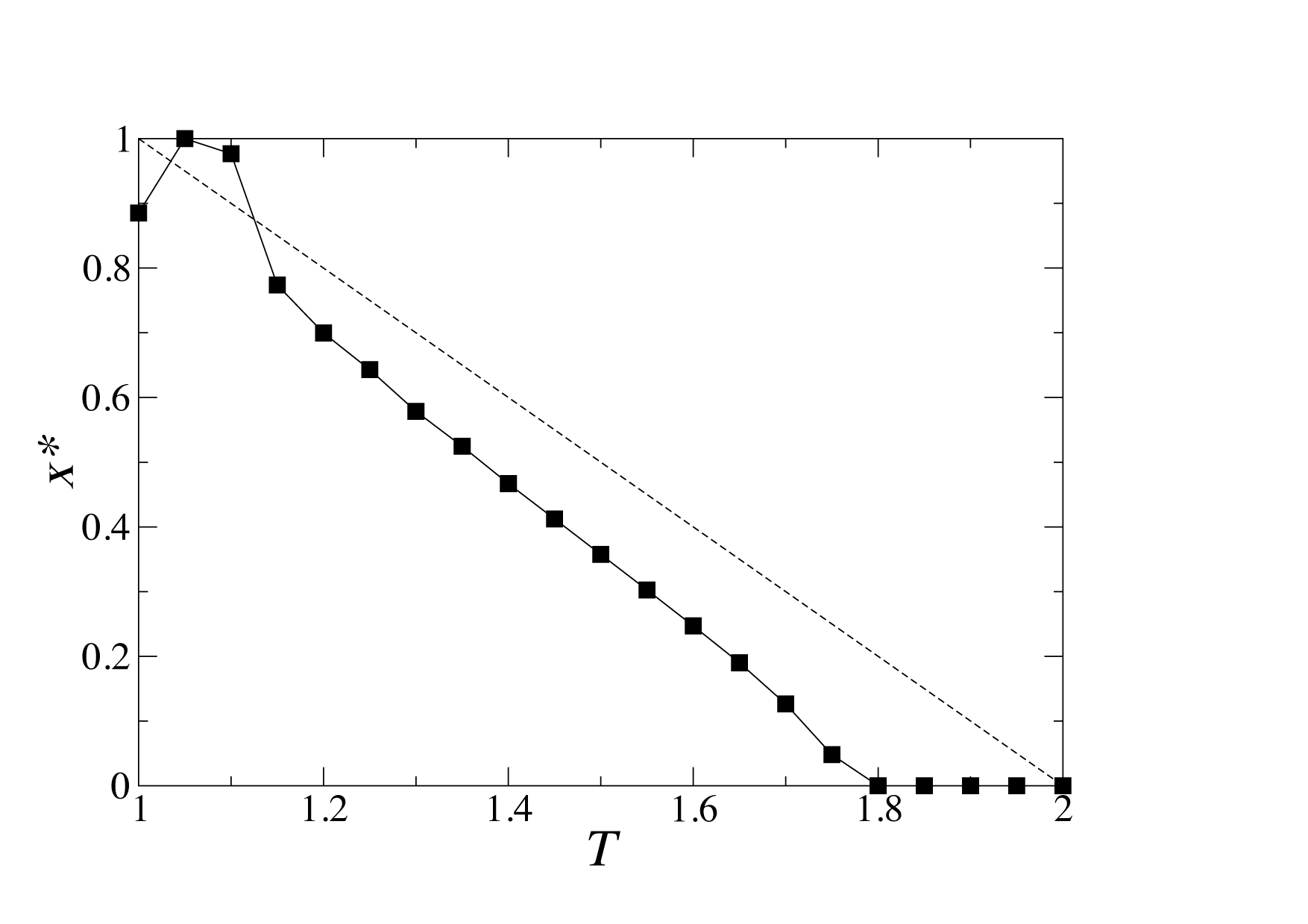}
\caption{Asymptotic density of cooperators $x^*$ in a square lattice with degree
$k=8$, when the game is the Snowdrift defined by Eq.~(\ref{eq:hauert-sd}), as
studied by Hauert and Doebeli \cite{hauert:2004} (update rule: replicator rule). Lines are a guide to the
eye. The result for a well mixed population is displayed as reference as a
dashed line.}
\label{fig:hauert}
\end{figure}

In 1996 Killingback and Doebeli published a paper \cite{killingback:1996} that
studied a spatial model similar to that of \cite{nowak:1992}, but
they considered Hawk-Dove games (equivalent to Snowdrift games) instead
of Prisoner's Dilemmas. They reported a lowering of the proportion of Hawks
in the population, which in terms of the evolution of cooperation means a
promotion of the cooperative strategy. Later, Hauert and Doebeli published
another
result \cite{hauert:2004} reporting an inhibition of cooperation precisely on
Snowdrift games in a spatial model. They studied the following parametrization
of Snowdrift games
\begin{equation}
\label{eq:hauert-sd}
\begin{array}{cc}
  & \begin{array} {cc} \mbox{C} & \mbox{D} \end{array} \\
  \begin{array}{c} \mbox{C} \\ \mbox{D} \end{array} &
  \left( \begin{array}{cc} 1 & 2-T \\ T & 0 \end{array} \right),
\end{array}
\end{equation}
again with $1 \le T \le 2$. The unexpected result obtained by the authors is
displayed in Fig.~\ref{fig:hauert}. Only for low $T$ there is some improvement
in cooperation, whereas for medium and high $T$ cooperation is inhibited. This
is a surprising result, considering \cite{killingback:1996} and because the
basic game, Snowdrift, is in principle more favorable to cooperation. Its only
stable equilibrium is a mixed strategy population with some density of
cooperators (\ref{eq:mixedeq}), whereas the unique equilibrium in Prisoner's
Dilemma is full defection (see Fig.~\ref{fig:compnet}). The authors explained
their result in terms of the inhibition of cluster formation and growth, at
the
microscopic level, caused by the payoff structure of Snowdrift games. We find
this explanation controversial, because the results of Nowak and May were
obtained precisely near the boundary between Prisoner's Dilemma and Snowdrift,
so this argument implied a discontinuous transition in the microscopic dynamics
at this boundary. Nevertheless, it is very easy to check that there is not a
discontinuity neither in the payoff matrix nor in the equilibrium structure of
the games at this boundary. Hence, where does this transition in the microscopic
dynamics come from?

\begin{figure}
\centering
\includegraphics[width=7.5cm]{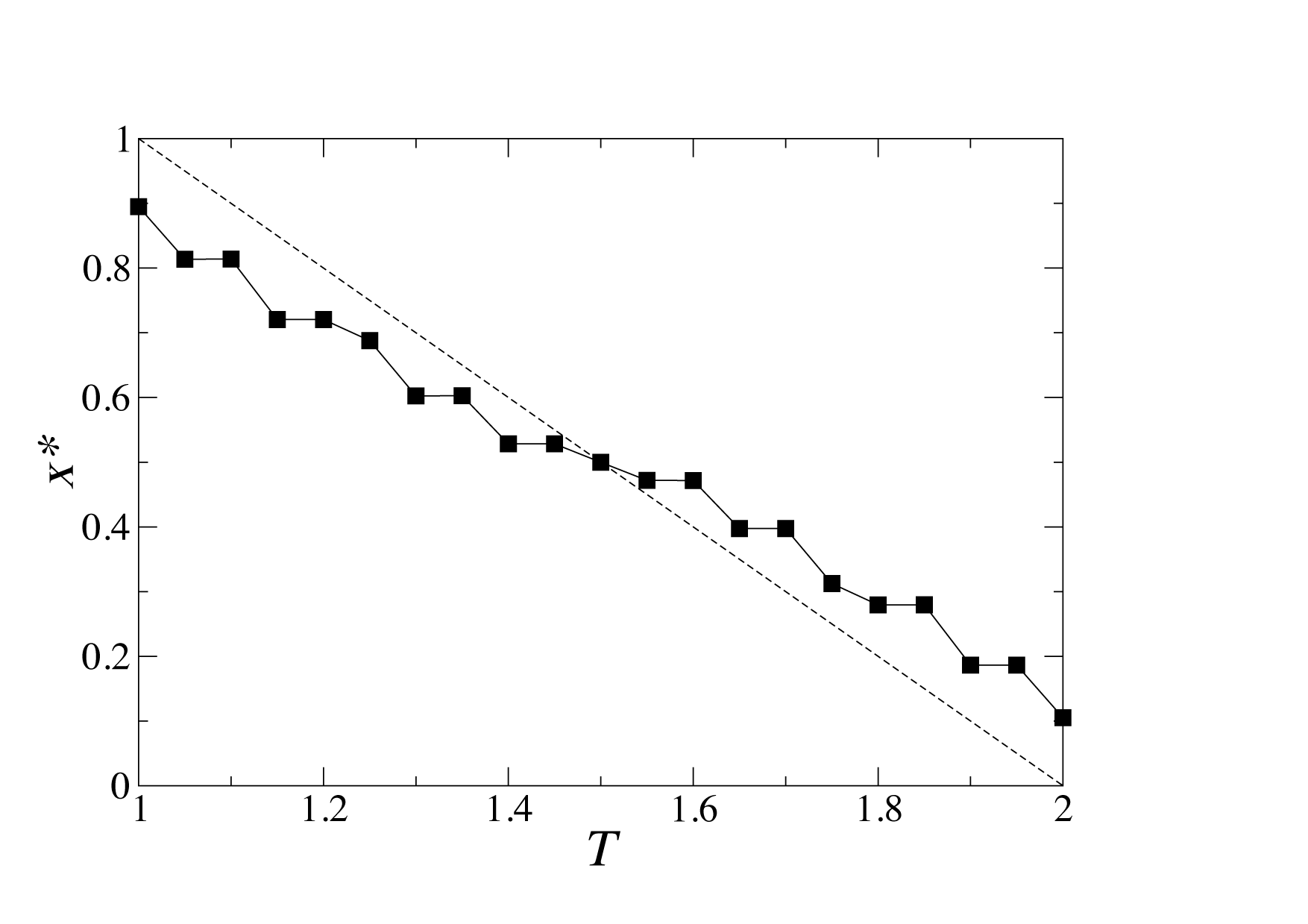}
\caption{Asymptotic density of cooperators $x^*$ in a square lattice with degree
$k=8$, when the game is the Snowdrift defined by Eq.~(\ref{eq:hauert-sd}), as
studied by Sysi-Aho and co-workers \cite{sysi-aho:2005} (update rule: best response). Lines are a guide to
the eye. The result for a well mixed population is displayed as reference as a dashed line. Note how the promotion or inhibition of cooperation does not
follow the same variation as a function of $T$ that in the case studied by
Hauert and Doebeli \cite{hauert:2004} (Fig.~\ref{fig:hauert}).}
\label{fig:sysi-aho}
\end{figure}

Another result concerning Snowdrift games was published in 2005 by Sysi-Aho and
co-workers \cite{sysi-aho:2005}. They studied the same one-dimensional
parametrization of Snowdrift games (\ref{eq:hauert-sd}), but used \emph{myopic
best response} \cite{matsui:1992,blume:1993} as the rule for the update of
strategies. They reported a modification in the cooperator density at
equilibrium, with an increase for some subrange of the parameter $T$ and a
decrease for the other, as Fig.~\ref{fig:sysi-aho} shows. It is very
remarkable that the effect on cooperation (promotion or inhibition) is in this
case opposite to the result of Hauert and Doebeli \cite{hauert:2004}
(Fig.~\ref{fig:hauert}).

Following those works on Snowdrift or Hawk-Dove games on spatial lattices,
Tomassini and co-workers \cite{tomassini:2006} performed an exhaustive study
with an equivalent parametrization of Hawk-Dove games, considering several
update rules and also other topologies as random and small-world networks. They
concluded that the influence of spatial structure on cooperation with this kind
of games can be positive or negative, and that it depends on the update rule, on
the value of the parameter $T$ and, to a lesser extent, on the synchronicity or
asynchronicity of the updating. An important question is pending here therefore:
Is this sensitivity to the update rule an exclusive feature of Snowdrift games
or, on the contrary, should it be expected to occur with other games, such as
the Prisoner's Dilemma?

Another open question in the existing literature is the importance of
the \emph{transitivity of links} or \emph{clustering} \cite{newman:2003} in
the influence of spatial structure on the evolution of cooperation. In an
influential paper published by Cohen and co-workers \cite{cohen:2001}, the
authors proved that the positive effects on cooperation of some regular
lattices were in fact attainable with random networks of the same degree. They
concluded that, for the model networks they were studying, the relevant
topological feature was not the spatial arrangement of links, and their
subsequent correlations or clustering, but the \emph{context preservation} of
players during the iterated game, already present in a random network. Later,
Ifti and co-workers \cite{ifti:2004} performed a more exhaustive study in the
space of network models, considering different topologies and several degrees,
reaching the opposite conclusion that ``the clustering is the factor that
facilitates and maintains high average investment values'' (i.e.\ cooperation).
The subsequent work of Tomassini and co-workers \cite{tomassini:2006} somehow
confirmed this point, as they showed, in the different context of Hawk-Dove
games, the key role of the network clustering in the positive or negative
influence on cooperation.

Finally, all these previous studies have considered the influence of spatial networks in the case of \emph{strong selection pressure}, which means that the fitness of individuals is totally determined by the payoffs that result from the game. In
general this may not be the case, and so to relax this restriction the fitness
can be expressed as $f = 1 - w + w \pi$ \cite{nowak:2004a}. The parameter $w$
represents the intensity of selection and can vary between $w=1$ (strong
selection limit) and $w \gtrsim 0$ (weak selection limit). The weak selection
limit has the nice property of been usually tractable in an analytic manner, and
thus it has become a key aspect of the field, with a great deal of related work being
done in the last few years. For instance, Ohtsuki and Nowak have studied
evolutionary games on homogeneous random networks using this approach
\cite{ohtsuki:2006}, finding an interesting relation with replicator dynamics on
well mixed populations. Using our normalization of the game
(\ref{eq:payoff-matrix}), their result can be written as the following
payoff matrix
\begin{equation}
\left(
  \begin{array}{cc} 1 & S + \Delta \\ T - \Delta & 0 \end{array}
\right).
\end{equation}
This means that the evolution in a population structured according to a random
homogeneous network, in the weak selection limit, is the same as that of a well
mixed population with a game defined by this modified payoff matrix. The effect
of the network thus reduces to the term $\Delta$, which depends on the
network degree, the game and the update rule. With respect to the influence of
cooperation it admits a very straightforward interpretation: If both the
original and the modified payoff matrix correspond to a Harmony or Prisoner's
Dilemma game, then there is no influence, because the population ends
up equally in full cooperation or full defection; otherwise, cooperation is
enhanced if $\Delta > 0$, and inhibited if $\Delta < 0$. The actual values of
$\Delta$, for the update rules they studied, namely Pairwise Comparison (PC),
Imitation (IM) and Death-Birth (DB) (see \cite{ohtsuki:2006} for full details),
are \begin{eqnarray}
\Delta_{PC} &=& \frac {S - (T-1)} {k-2} \\
\Delta_{IM} &=& \frac {k + S - (T-1)} {(k+1)(k-2)} \\
\Delta_{DB} &=& \frac {k + 3[S - (T-1)]} {(k+3)(k-2)} ,
\end{eqnarray}
$k$ being the degree of the network. (Note that these results correspond to the
weak selection limit.)

Considering the influence on Prisoner's Dilemma, first note that $S<0$ and
$T>1$ in all games of this kind, so the term $S-(T-1)$ is always negative.
It is easy to see that the most favorable rule for cooperation is
then Imitation, but even in this case the network has no effect on the outcome
(the game remains in the Prisoner's Dilemma square) if $S<-1/(k-1)$ and
$T>1+1/(k-1)$. For network degrees as the examples above (e.g.\ $k=8$ yields
$1/(k-1) \approx 0.143$) this means that the evolutionary outcome of
most games in the Prisoner's Dilemma square remains unaffected, i.e.\ most games end up in full defection anyway. With the
Death-Birth rule the region with no influence is even bigger ($S<-1/(k+1)$ and
$T>1+1/(k+1)$) and with Pairwise Comparison there is no effect on the
outcome of any Prisoner's Dilemma. Therefore, several questions arise here: Is
this reduced effect on cooperation caused by the weak selection pressure, by the
homogeneous random topology (which lacks spatial structure) or by the
combination of the two? What is the outcome in regular lattices with weak
selection?

\begin{figure*}
\centering
\includegraphics[width=0.99\textwidth]{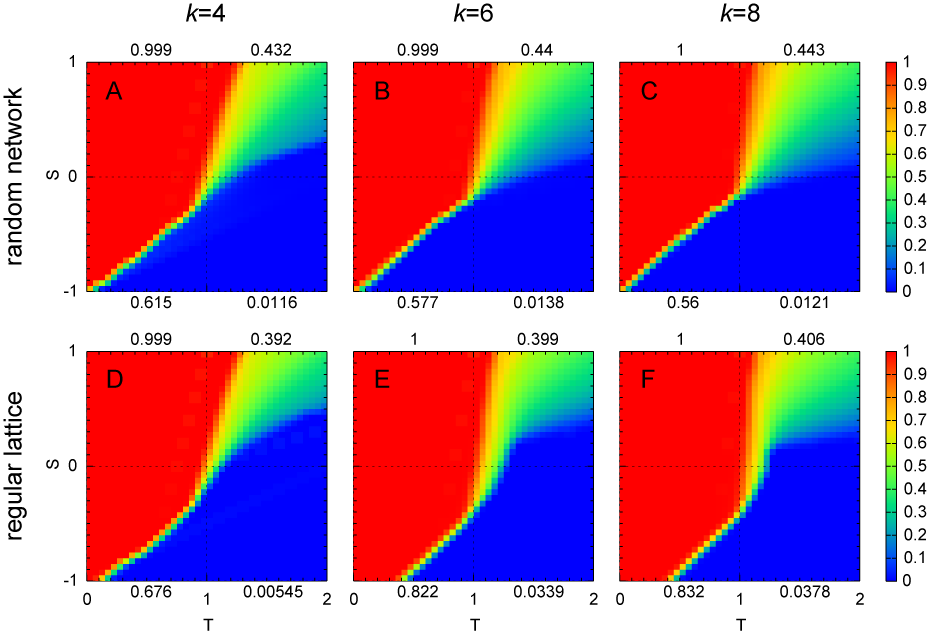}
\caption{(Color online) Asymptotic density of cooperators $x^*$ in homogeneous random networks
(upper row, A to C) compared to regular lattices (lower row, D to F), with
degrees $k=$ 4 (A, D), 6 (B, E) and 8 (C, F). The update rule is the replicator
rule and the initial density of cooperators is $x^0 =0.5$.
The values of the global index of cooperation in the upper row
display the weak effect of homogeneous random networks (compare with Fig.~\ref{fig:compnet}(b)), which is
opposite in Stag Hunt (lower-left square) and Snowdrift games (upper-right
square). Comparing both rows, the differences are specially significant in
Stag Hunt for $k=$ 6 and 8, revealing the strong promotion of cooperation
caused by regular lattices with large clustering (see text). The influence
on Harmony and Prisoner's Dilemma games is negligible in all cases.}
\label{fig:lattice-repl}
\end{figure*}

We thus see that there are indeed some important open questions
regarding the synchronicity of the updating, the choice of the update rule, the
clustering of the network and the intensity of selection. Previous research has
proved that these issues play a role in the effect of spatial structure on the
evolution of cooperation, for particular parametrizations of Prisoner's Dilemmas
and Snowdrift games. However, different conclusions were obtained
depending on the game, and as a consequence a comprehensive picture of these
issues on the wider space of $2 \times 2$ games, including other games as for
example coordination or Stag Hunt games, is not available. To provide it, we
designed a unified framework of simulation and performed an exhaustive and
systematic simulation study, covering all the possible configurations of games,
networks and update rules introduced above. The results that we have obtained
and, more importantly, the general conclusions that they have allowed us to
reach are presented in the remaining of the paper.

\section{A unified study of evolutionary games on spatial networks}

\label{sec:results}

\begin{figure*}
\centering
\includegraphics[width=0.99\textwidth]{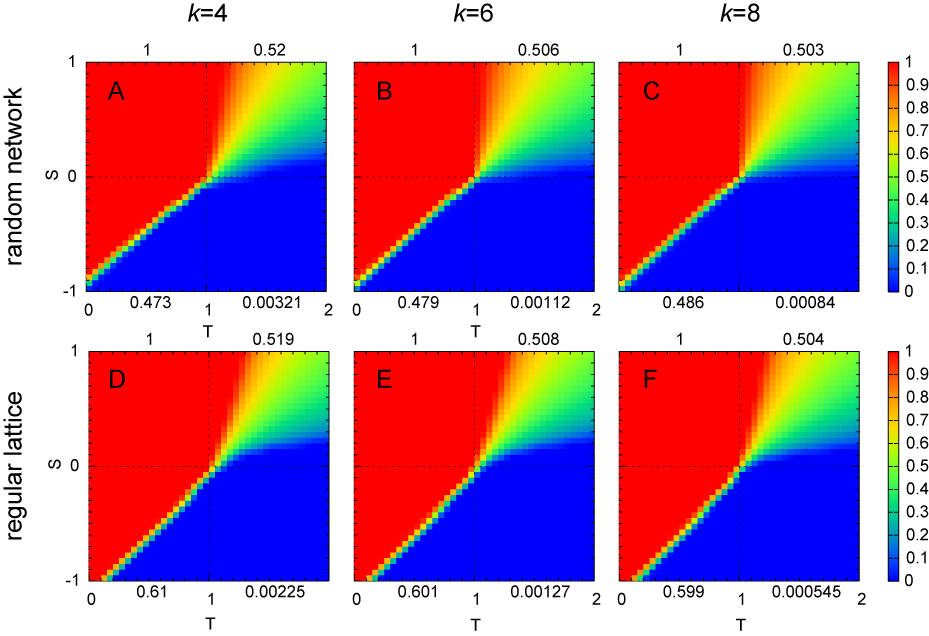}
\caption{(Color online) Asymptotic density of cooperators $x^*$ in homogeneous random networks
(A to C) compared to regular lattices (D to F), with degrees $k=$ 4 (A, D), 6
(B, E) and 8 (C, F). The update rule is the Moran rule and the initial
density
of cooperators is $x^0 =0.5$. With this update rule the effect is very small, in comparison with the other stochastic rules (see Fig.~\ref{fig:lattice-repl}). In any case, the most relevant effect, albeit much weaker than in the other cases, is again a promotion of cooperation in Stag Hunt games with regular lattices (D to F).}
\label{fig:lattice-moran}
\end{figure*}

\begin{figure*}
\centering
\includegraphics[width=0.99\textwidth]{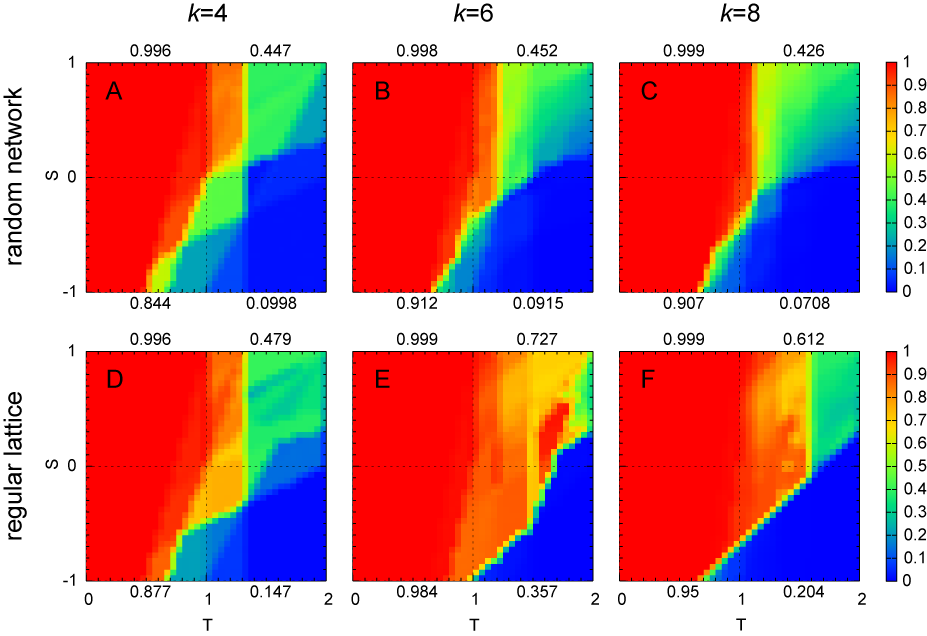}
\caption{(Color online) Asymptotic density of cooperators $x^*$ in homogeneous random networks
(upper row, A to C) compared to regular lattices (lower row, D to F), with
degrees $k=$ 4 (A, D), 6 (B, E) and 8 (C, F). The update rule is unconditional
imitation and the initial density of cooperators is $x^0 =0.5$.
Cooperation is fostered in Stag Hunt games in all cases. As
in Fig.~\ref{fig:lattice-repl}, the differences between
homogeneous random networks and regular lattices of the same
degree are significant only when the regular lattices have large clustering
($k=$ 6 and 8). Note the non-monotonicity of the results, compared with the other update rules (see text). With this update rule the promotion of
cooperation clearly extends to Snowdrift and Prisoner's Dilemma games.}
\label{fig:lattice-uncimit}
\end{figure*}

To assess the effect of spatial structure we have studied the evolution of
the four classes of $2 \times 2$ games presented in Sec.~\ref{sec:evol-games} on
populations distributed over regular lattices of degrees $k=$ 4, 6, and 8, along
with different rules for the update of strategies. Additionally, to discern what
can be attributed to the spatial
distribution of links and what to the mere limitation in the number of
neighbors, we have compared in every case with the results on homogeneous
random networks of the same degree. To ensure a correct comparison of results
we have employed as much as possible the same parameters in all the
simulations (see the Appendix for full details).

Figure~\ref{fig:lattice-repl} displays the comparison of results between
homogeneous random networks and regular lattices, for the particular case of
the replicator update rule. A close examination of the different panels,
together with the reference outcome of Fig.~\ref{fig:compnet}(b), yields the
following conclusions. In the first place, the influence of homogeneous
networks in combination with the replicator rule is very weak in Harmony and
Prisoner's Dilemma games, as the mean indexes of cooperation that appear by
each square show. The effect is thus concentrated on Stag Hunt and Snowdrift
games, and it consists, in all cases, on a promotion of cooperation in Stag Hunt
and an inhibition in Snowdrift. Considering the dependence on network degree,
the influence of homogeneous random networks diminishes for larger degrees, as
can be seen in the graphs A-C of Fig.~\ref{fig:lattice-repl}, from left
to right. This variation with degree is, however, very different in the case of
regular lattices. Whereas for $k=4$ the outcome is similar for both kinds of
networks --albeit slightly stronger in the lattices--, the influence of the
network increases
significantly for lattices of larger degrees. These are precisely
the conditions where the influence of spatial structure shows up, as a large
promotion of cooperation in Stag Hunt and a comparatively smaller inhibition
in Snowdrift. We will justify in the next section that the key topological
feature underlying this effect of spatial structure is the presence of
\emph{clustering}
in the network, understood as link transitivity or, equivalently, triangles in
the graph \cite{newman:2003}. For our purposes the clustering coefficient
$\mathcal{C}$ of a network is defined as the probability that any two
neighbors of a given node are neighbors themselves. Thus, regular lattices of degree $k=4$ have $\mathcal{C}=0$, whereas those of degrees $k=$ 6 and 8 have,
respectively, values of $\mathcal{C}=2/5=0.4$ and $\mathcal{C}=3/7 \approx
0.43$.

Changing the rule for the update of strategies to the multiple replicator rule
only causes small quantitative changes in the results. The other stochastic
rule, i.e.\ the Moran rule, has a larger impact, as it clearly reduces the
influence of the network, but it maintains to some extent the positive effect on
Stag Hunt of spatial lattices, as can be seen in Fig.~\ref{fig:lattice-moran}.

The rule, however, that has the greatest impact on the evolutionary outcome is
unconditional imitation. Figure~\ref{fig:lattice-uncimit} presents the
results for this rule. In the first place, homogeneous random
networks clearly have an influence on the evolutionary outcome with this rule,
for all the degrees considered. They clearly promote cooperation in Stag Hunt,
and even some influence is noticeable on Prisoner's Dilemma.
Second, it is very remarkable the similarity in results between random
networks and regular lattices for degree $k=4$, with the transition lines
located at the same positions (see \cite{schweitzer:2002} for a detailed study of these transition lines on the regular lattice). And finally, large differences appear again when the spatial lattices have clustering, i.e.\ for degrees $k=$ 6 and 8. In these cases spatial structure not only induces almost full cooperation in Stag Hunt games, but also enforces cooperation notably in Snowdrift and Prisoner's Dilemma games, to the greatest extent we have obtained for any setting in the systematic study we present in this paper.

It is also interesting to note, in the case of unconditional imitation compared to the other update rules, the non-monotonous variation of results across the $ST$-space. With this rule the evolutionary outcome is determined by the dominance or coexistence of certain privileged configurations of cooperators and/or defectors (see Sec.~\ref{sec:discussion} and also \cite{schweitzer:2002}). This dominance or coexistence depends in turn on the balance of payoffs along the corresponding interfaces between configurations. The balance, i.e.\ which configuration obtains more payoff than some other, switches when crossing some boundary on the $ST$-space, giving rise to a non-monotonous response as the leading configurations change when varying the parameters $S$ and $T$ (see Figs.~\ref{fig:nowak} and~\ref{fig:lattice-uncimit}).

From the viewpoint of the open questions posed in
Sec.~\ref{sec:previous-results}, these results prove that the effect of spatial
structure, in the context of $2 \times 2$ games, is highly dependent on the
update rule. This dependence explains the apparent contradiction between the
promotion of cooperation reported by Nowak and May for Prisoner's Dilemmas
\cite{nowak:1992} and by Killingback and Doebeli for Snowdrift games
\cite{killingback:1996} vs. the inhibition on the same Snowdrift games reported
by Hauert and Doebeli \cite{hauert:2004}. In the first two works the update rule
was unconditional imitation, whereas in the third the authors used the
replicator rule. Note that this explanation is in agreement with the discussion
in \cite{tomassini:2006}, which deals exclusively with Snowdrift games.
Comparing the Figs.~\ref{fig:lattice-repl}~F
and~\ref{fig:lattice-uncimit}~F, it is clear that the influence on games around the boundary between Prisoner's Dilemma and Snowdrift is similar and without discontinuities if the update rule is the same.

In what concerns the Prisoner's Dilemma, the above results also prove that the
promotion of cooperation in this game is not robust against changes in the
update rule, because the beneficial effect of spatial lattices practically 
disappears for rules different from unconditional imitation, when seen in the
wider scope of the $ST$-plane. Notice that this conclusion supersedes previously
published work, as for example \cite{hauert:2002}, where it was claimed that
often spatial extension was indeed capable of promoting cooperative behavior,
in particular for the Prisoner's Dilemma for a small but important parameter
range. Further, \cite{hauert:2002} stated that the conclusions were
robust and appeared to be almost independent of the update rule of the lattice.
As we have explained in the introduction, there were problems with the
simulations in \cite{hauert:2002}. 

On the contrary, coordination or Stag Hunt games are the games where the
positive effect on cooperation of spatial structure is robust against changes in
the update rule, in particular the introduction of stochasticity. This property,
together with the fact that this kind of games can be an attractor on
evolutionary dynamics of the game payoffs themselves \cite{fort:2008}, may be of
a special relevance in the problem of the evolution of cooperation.

Another important conclusion that can be drawn from the results presented so far
is the relevance of the \emph{clustering coefficient} to the effect of spatial
structure: only when the clustering coefficient is high, the spatial
distribution of the population makes a difference in comparison to a random
arrangement of links that just limits the number of interactions of players and
preserves their context. This difference holds in the $ST$-plane, i.e.\ for
whichever game among those considered, and is robust against changes in the
update rule. This point explains the difference between the conclusions of Cohen
\emph{et al.\ } \cite{cohen:2001} and those of Ifti \emph{et al.\ }
\cite{ifti:2004} and Tomassini \emph{et al.\ } \cite{tomassini:2006}. In
\cite{cohen:2001}, rectangular lattices of degree $k=4$ were considered, which
have strictly zero clustering because there are not closed triangles in the
network, hence finding no differences in outcome between the spatial and the
random topology. In the latter case, on the contrary, both studies employed
rectangular lattices of degree $k=8$, which do have clustering, and thus they
identified it as a key feature of the network, for the
particular parametrizations of the games they were studying, namely Prisoner's
Dilemma \cite{ifti:2004} and Snowdrift \cite{tomassini:2006}.

\begin{figure*}
\centering
\includegraphics[width=11cm]{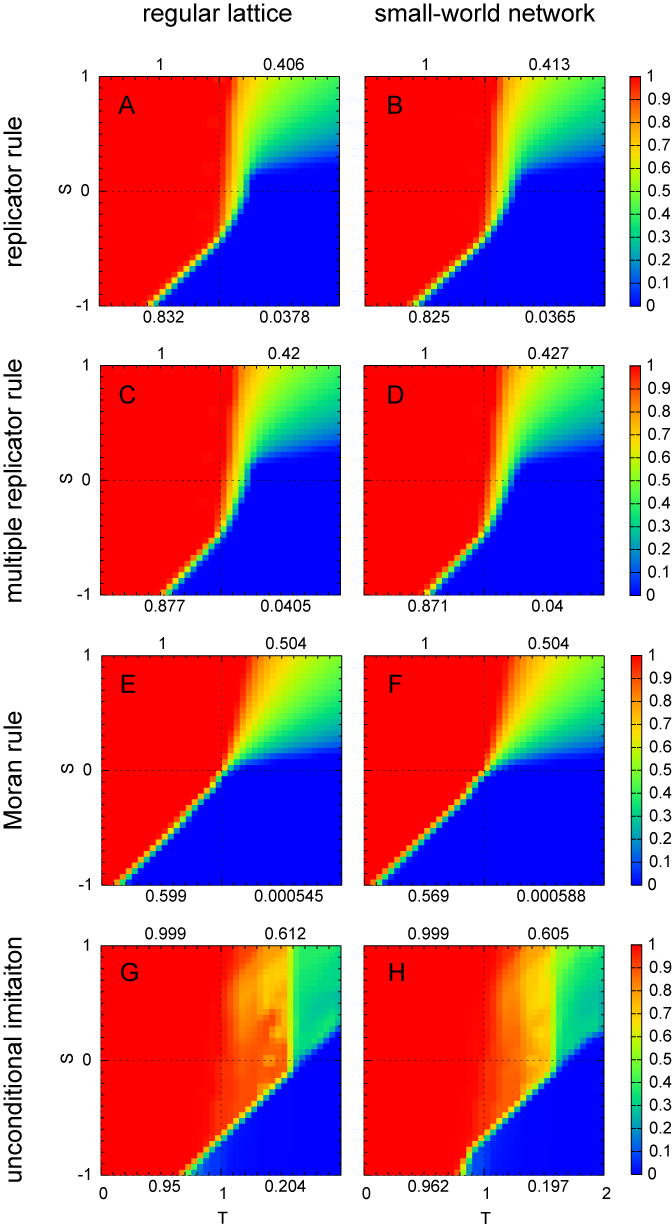}
\caption{(Color online) Asymptotic density of cooperators $x^*$ in regular lattices (left
column) compared to small-world networks (right column), all with degree $k=8$. The
update rules are: replicator rule (first row), multiple replicator rule (second row),  Moran rule (third row) and unconditional imitation (fourth row). The initial density of cooperators is $x^0 =0.5$. Notice that the panels A, E and G are the same as, respectively,
the panels~\ref{fig:lattice-repl}~F, \ref{fig:lattice-moran}~F and \ref{fig:lattice-uncimit}~F, repeated
here to facilitate the comparison. The evolutionary outcomes are practically identical for all the update rules, showing that the effect of small-world networks on the asymptotic state is due to the high clustering, also present in the regular lattices used to generate them.
The probability of reshuffling used to generate the small-world networks was
$p=0.01$ (see text).}
\label{fig:small-world}
\end{figure*}

\begin{figure*}
\centering
\includegraphics[width=11cm]{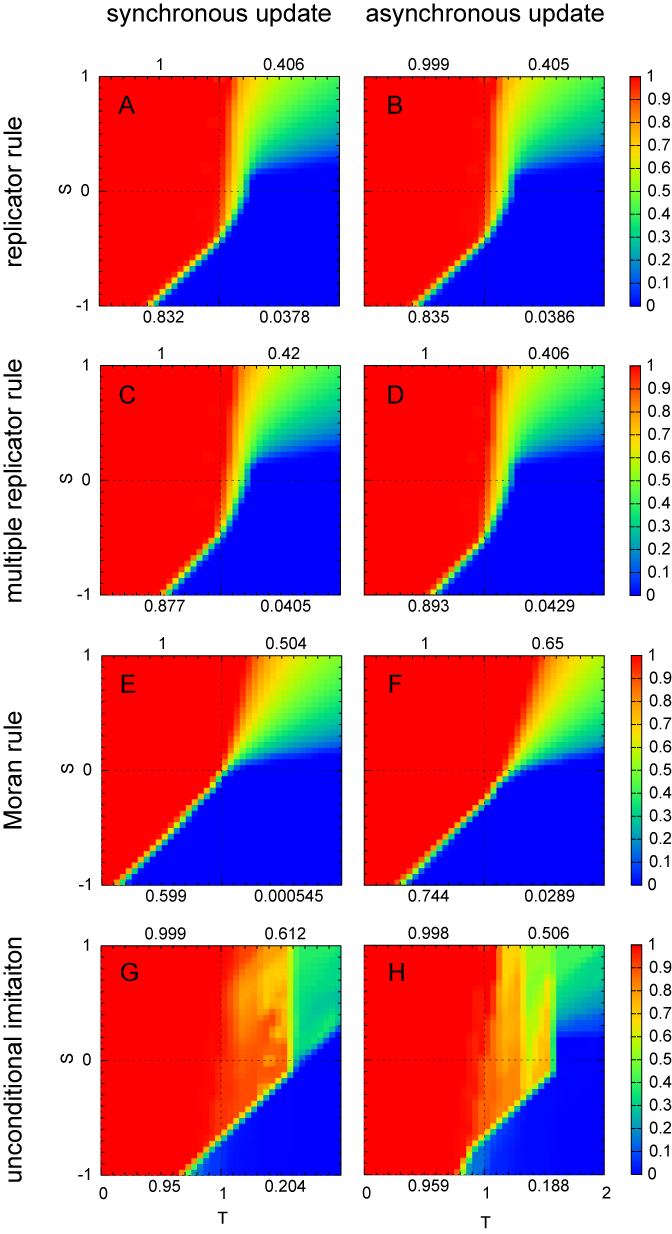}
\caption{(Color online) Asymptotic density of cooperators $x^*$ in regular lattices of
degree $k=8$, with synchronous update (left column) compared to asynchronous
(right column). The update rules are: replicator rule (first row), multiple replicator rule (second row), Moran rule (third row) and 
unconditional imitation (fourth row). The initial density of cooperators is $x^0=0.5$. As in Fig.~\ref{fig:small-world}, the panels A, E and G are the same as, respectively, the panels~\ref{fig:lattice-repl}~F, \ref{fig:lattice-moran}~F and \ref{fig:lattice-uncimit}~F. With the replicator and multiple replicator rules the evolutionary outcome is very similar, whereas some differences appear in the case of the Moran rule. With unconditional imitation the results are also quite similar, but there are
differences for some points, specially those in the Snowdrift square with $S
\lesssim 0.3$ and
$T>5/3 \approx 1.67$. The particular game studied by \cite{huberman:1993},
which reported an inhibition of cooperation due to the asynchronous update,
belongs to this region.}
\label{fig:async-1}
\end{figure*}

\begin{figure*}
\centering
\includegraphics[width=0.99\textwidth]{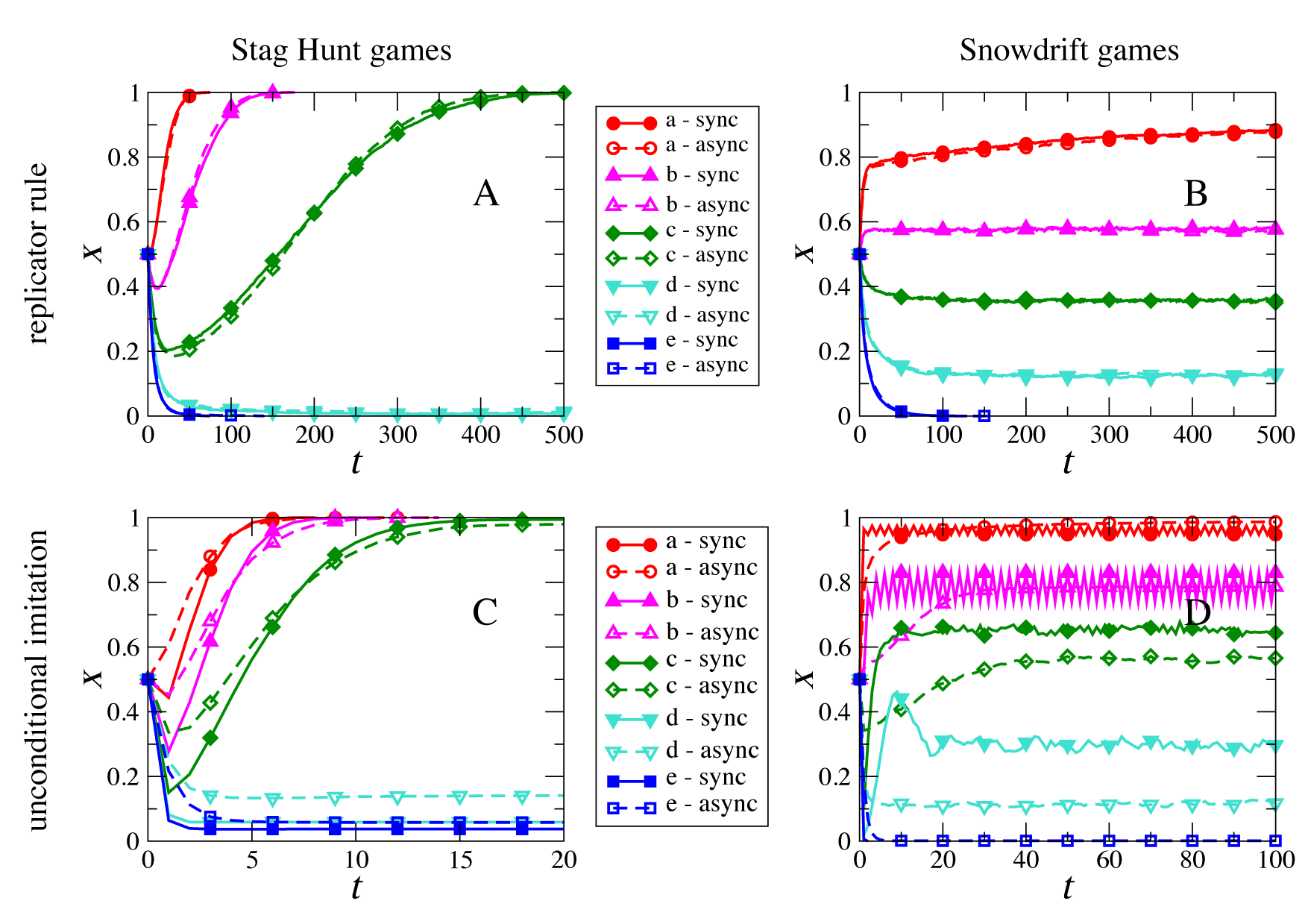}
\caption{(Color online) Time evolution of the density of cooperators $x$ in regular lattices
of degree $k=8$, for typical realizations of Stag Hunt (left column) and
Snowdrift games (right
column). The update rules are the replicator rule (upper row) and
unconditional imitation (lower row). The realizations with synchronous update are represented by continuous lines and filled symbols, whereas the asynchronous ones appear with dashed lines and empty symbols. Colors and symbols are as indicated in the legends. The Stag Hunt games for the replicator rule (A) are: a, $S=-0.4$, $T=0.4$; b, $S=-0.5$, $T=0.5$; c, $S=-0.6$, $T=0.6$; d,
$S=-0.7$, $T=0.7$; e, $S=-0.8$, $T=0.8$. For unconditional imitation the Stag
Hunt games (C) are: a, $S=-0.6$, $T=0.6$; b, $S=-0.7$, $T=0.7$; c, $S=-0.8$,
$T=0.8$;
d, $S=-0.9$, $T=0.9$; e, $S=-1.0$, $T=1.0$. The Snowdrift games are, for both
update
rules (B, D): a, $S=0.9$, $T=1.1$; b, $S=0.7$, $T=1.3$; c, $S=0.5$, $T=1.5$; d,
$S=0.3$, $T=1.7$; e, $S=0.1$, $T=1.9$. The initial density of cooperators is
$x^0 =0.5$ in all cases. The time scale of the asynchronous realizations has
been re-scaled by the size of the population $N = 10^4$, so that for both kinds of update a time step represents the same number of update events in the
population. Figures A and B show that, in the case of the stochastic rule, not
only the outcome but also the time evolution is independent of the kind of
update. With unconditional imitation the results are also very similar for Stag Hunt (C), but not so much in Snowdrift (D), displaying the influence of the type of updating in this case. Notice also that unconditional imitation causes a much faster evolution than the replicator rule.}
\label{fig:async-2}
\end{figure*}

\begin{figure*}
\centering
\includegraphics[width=0.99\textwidth]{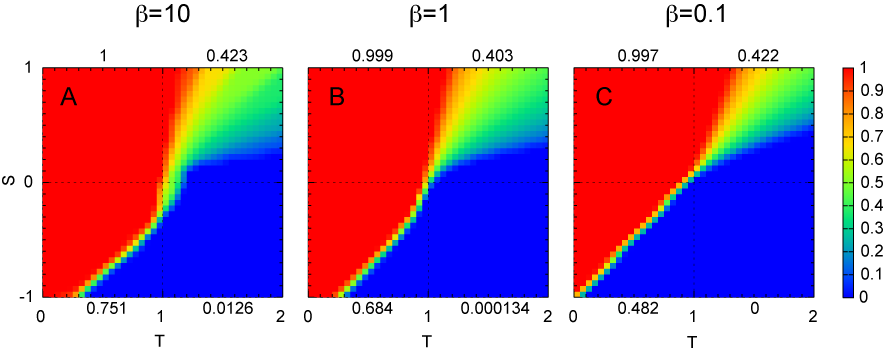}
\caption{(Color online) Asymptotic density of cooperators $x^*$ in regular lattices of degree
$k=8$, for the Fermi update rule with $\beta$ equal to 10 (A), 1 (B) and 0.1
(C). The initial density of cooperators is $x^0=0.5$. For high $\beta$ the result is quite similar to that obtained with the replicator rule
(Fig.~\ref{fig:lattice-repl}~F). As $\beta$ decreases, or equivalently for
weaker intensities of selection, the influence becomes smaller and more
symmetrical between Stag Hunt and Snowdrift games.}
\label{fig:lattice-fermi}
\end{figure*}

To check the robustness of these conclusions, and also for the intrinsic interest of complex topologies, we have included small-world networks \cite{newman:2003,boccaletti:2006} in our systematic simulations. Our
procedure to build the network is based on the Watts-Strogatz algorithm \cite{watts:1998}. We start from a regular lattice and perform, with low probability, a random reshuffling of links preserving the degree of the nodes, with the aim of lowering the mean distance between nodes while maintaining
the high clustering coefficient and the homogeneity in degree. In
our case, we have started from the two-dimensional lattices presented above, so we can consider the resulting network as a slightly disordered lattice, which maintains, however, the local property of large clustering. Figure~\ref{fig:small-world} shows a
comparison between the results obtained with the small-world topology and the
corresponding initial regular lattice, for all four update rules. The
evolutionary outcomes are almost identical, and the tiny quantitative
differences can only be noticed by means of the mean cooperation
index associated with each game. Consequently, a small number of defects in the
spatial structure of a lattice does not alter its effect on the evolutionary
outcome, even if they lead to a drastic decrease of the diameter of the
network. Very similar results are obtained for other values of the probability
of reshuffling, as long as the network clustering is preserved.

Considering these results from another point of view, we can state that the
influence on cooperation of the small-world topology, in what concerns the
evolutionary outcome and at least for the kinds of games and update rules
considered in this work, is determined by its feature of high clustering, which
in the Watts-Strogatz model comes from the regular lattice employed to generate
the network. This conclusion is in agreement with existent theoretical work on
Prisoner's Dilemma \cite{abramson:2001,masuda:2003,tomochi:2004} and its
extensions \cite{wu:2005,wu:2006}, on Snowdrift games \cite{tomassini:2006}, and
also with experimental studies on coordination games \cite{cassar:2007}. The
reader should note that some of these works have reported a greater
\emph{efficiency} of small-world networks compared to regular lattices,
concerning the time of convergence to the stationary state, a property that we
have also verified in the broader scope of games of our study. It is reasonable
to conjecture that this improvement in efficiency is caused by the other typical
small-world feature, the low mean distance, as it is discussed in 
Sec.~\ref{sec:discussion}, which deals with the model dynamics.

All the results presented so far have been obtained with a synchronous update of
strategies. Recalling the impact of using asynchronous update
reported by Huberman and Glance \cite{huberman:1993}, we have studied all the previous models in its asynchronous version. We have found that the influence
of asynchronicity is the exception rather than the rule, and that this influence is very dependent on the update rule used. As Fig.~\ref{fig:async-1} shows, with the replicator and multiple replicator rules the evolutionary outcome is very similar, whereas for the Moran rule and unconditional imitation some differences appear. With this last rule, which was the one used in \cite{huberman:1993}, the only important variation takes place in a particular subset of Snowdrift games. Therefore, the discrepancy reported by \cite{huberman:1993} is restricted to a small subset of $2 \times 2$ games and the use of unconditional imitation as update rule. This conclusion is in agreement with previous work which has studied this issue in more limited settings, for the Prisoner's Dilemma \cite{nowak:1994,lindgren:1994,kirchkamp:2000} or the Snowdrift game \cite{kun:2006,tomassini:2006}. We can add that, if the time is re-scaled so that a time step represents the same number of update events in the population, then the time evolution is also very similar, specially for stochastic update rules, as Fig.~\ref{fig:async-2} reveals.

Finally, we have investigated the influence of the intensity of selection on
the evolutionary outcome of games in spatial lattices, by means of the Fermi
rule. Figure~\ref{fig:lattice-fermi} displays an example of the results for different intensities of selection, showing that weak selection has two important effects, in comparison with strong selection. First, it reduces the influence of the network on the evolutionary outcome, and second, it symmetrises its effects on Stag Hunt and Snowdrift games. These two effects
are consistent with the analytical calculations for the weak selection limit and homogeneous random networks \cite{ohtsuki:2006} discussed in the previous section. Our results
provide thus evidence of the following two key aspects of weak selection,
compared to strong selection, which to the best of our knowledge have not been reported in the literature. On one hand, the magnitude of the effect of
spatial structure is clearly smaller, with evolutionary outcomes much more
similar to those of an unstructured population. On the other hand, the nature
of the effect is also different, as weak selection has the same influence on
cooperation (promotion or inhibition) in Stag Hunt and Snowdrift games, namely
on those games symmetric with respect to the line $S+T=1$ in the $ST$-plane.

\section{Discussion}

\label{sec:discussion}

In the previous section we have given our answer, by means of a
systematic and exhaustive simulation program, to the open questions that we presented in Sec.~\ref{sec:previous-results}. We have highlighted the importance of the update rule and the singularity of unconditional imitation, which is the only rule among those studied here that yields a significant promotion of cooperation in the Prisoner's Dilemma. We have provided compelling evidence that, in general, the game in which the positive effect of spatial structure on cooperation is robust against changes in the update rule is the Stag Hunt, a coordination game. With stochastic rules, such as the replicator rule, this promotion of cooperation in Stag Hunt games is accompanied by an asymmetric influence, i.e.\ an inhibition of cooperation, in Snowdrift games. In addition, we have found that the strength of the effect of spatial structure is directly linked to the presence of clustering in the network. And finally, we have seen how weak selection attenuates the influence of spatial structure and symmetrizes the effects on Stag Hunt and Snowdrift games.

In the present Section we want to consider the causes of such features, looking for explanations rooted on the basic mechanisms that take place during the
evolution of the population. To this aim, let us consider a population with no
structure, i.e.\ connected by a complete network. A cooperator and a defector
obtain the following payoffs
\begin{eqnarray}
\pi_c &=& (n_c-1) + n_d S \approx N \left( x + (1-x)S \right) ,\\
\pi_d &=& n_c T = N x T ,
\end{eqnarray}
$N$ being the population size, $n_c$ and $n_d$ the total number of
cooperators and defectors, and $x$ the global fraction of cooperators.

With a structured population, however, each individual only plays
with her neighbors. Then, the payoffs are
\begin{eqnarray}
\pi_c &=& \hat{n}_c + \hat{n}_d S =
  k \left( \hat{x} + (1-\hat{x})S \right) ,\\
\pi_d &=& \hat{n}_c T = k \hat{x} T ,
\end{eqnarray}
$\hat{n}_c$ and $\hat{n}_d$ being the number of cooperators and defectors
that the player is connected to, and $\hat{x}$ the local fraction of
cooperators in the player's neighborhood. Note that $x$ is a global
variable, whereas $\hat{x}$ is defined for every player.
As a result, the effect of population structure can be understood
as the replacement of the global density $x$ by the player-dependent
local densities $\hat{x}$.

Let us now assume that the effect of spatial structure is
to increase the local densities $\hat{x}$ with respect
to the global density $x$. Considering Stag Hunt games in the first place, for a given initial condition $x^0$ there must be a subregion of the Stag Hunt
square in which $x_e$ verifies $x^0 < x_e < \hat{x}^0$. For these
games a complete network would produce an outcome of $x^* = 0$,
whereas the structured population would yield $x^* = 1$, with the
subsequent promotion of cooperation. On the other hand, for Snowdrift games, the equilibrium will be reached when $\hat{x}^*=x_e$. Since $x^* < \hat{x}^*$
this causes a global inhibition of cooperation.

This mechanism would explain the opposite effects on Stag Hunt and Snowdrift
games, and the absence of effects when the game has only one equilibrium, which
is the case with Harmony and Prisoner's Dilemma games. In fact, the increase in
the local densities is enforced by the correlations that arise as a result of
the spatial structure, i.e.\ because several neighbors observe simultaneously
the same high or low $\hat{x}$, as we will see below. For homogeneous random
networks and lattices with low clustering, correlations are weak, and hence
their influence on cooperation is hardly noticeable. Lattices with large
clustering, however, allow strong correlations to develop, raising the local
densities to such an extent that they have an important influence on the
evolutionary outcome. Considering the time evolution in the case of Stag Hunt
games, the local densities fluctuate over the population in the initial random
condition, with cooperators more or less connected to other cooperators. Those
with small $\hat{x}$ eventually disappear, while those with large $\hat{x}$
convert, with high probability, their defective neighbors to cooperators. This
is the point when the large clustering plays its crucial role: newly converted
cooperators will be connected not only to the cooperator whose strategy they
have just adopted, but also to some of her neighbors (because of the network
clustering), which are, with high probability, cooperators as well (because of
the high $\hat{x}$ of the initial cooperator). Hence the new cooperator will
also have a large local density of cooperators. Then, this process continues until the population reaches full cooperation.

\begin{figure*}
\centering
\includegraphics[width=13cm]{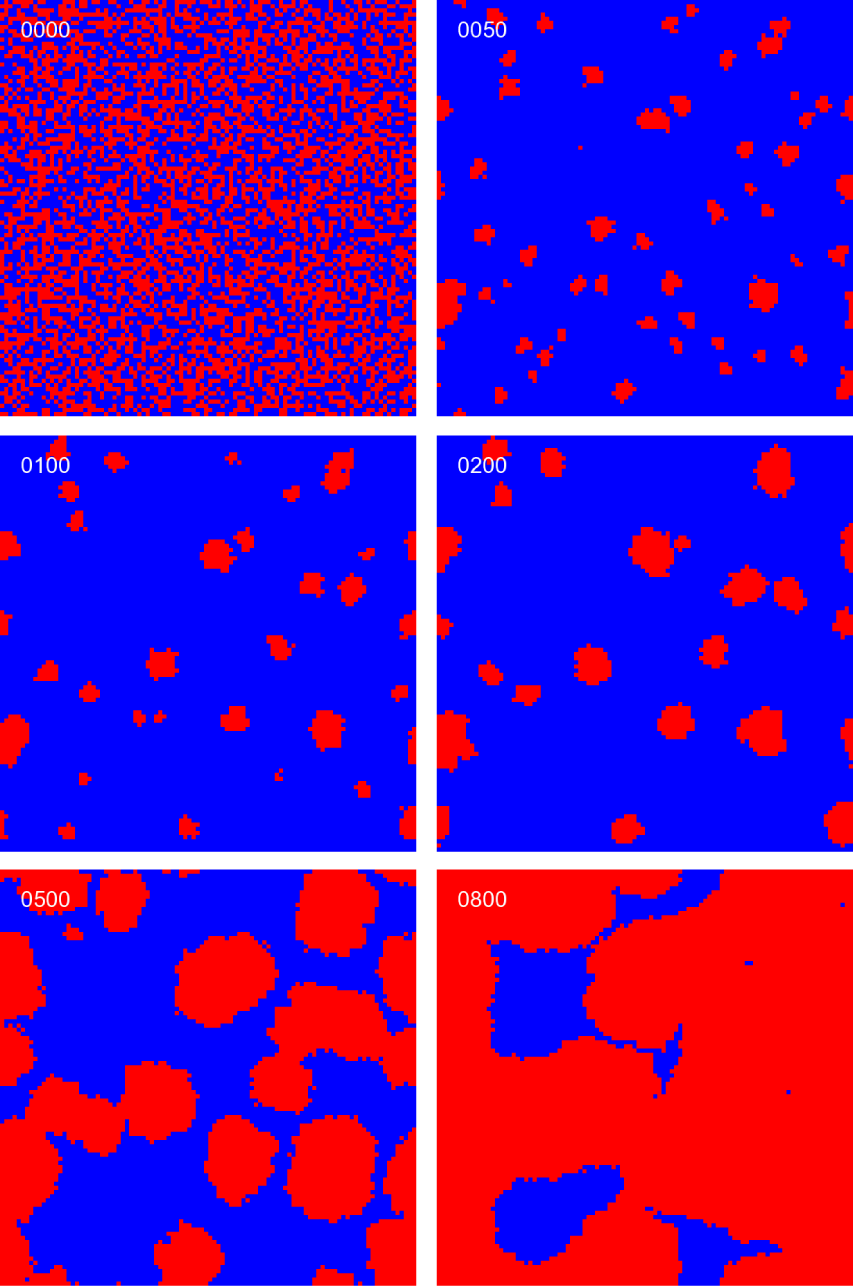}
\caption{(Color online) Snapshots of the evolution of a population on a regular lattice of
degree $k=8$, playing a Stag Hunt game ($S=-0.65$ and $T=0.65$). Cooperators are
displayed in red (light gray) and defectors in blue (dark gray). The update rule is the replicator rule
and the initial density of cooperators is $x^0 =0.5$. The upper left label shows
the time step $t$. During the initial steps, cooperators with low local
density of cooperators $\hat{x}$ disappear, meanwhile those with high local
density grow into the clusters that eventually take up the full population.}
\label{fig:evol-repl}
\end{figure*}

\begin{figure*}
\centering
\includegraphics[width=13cm]{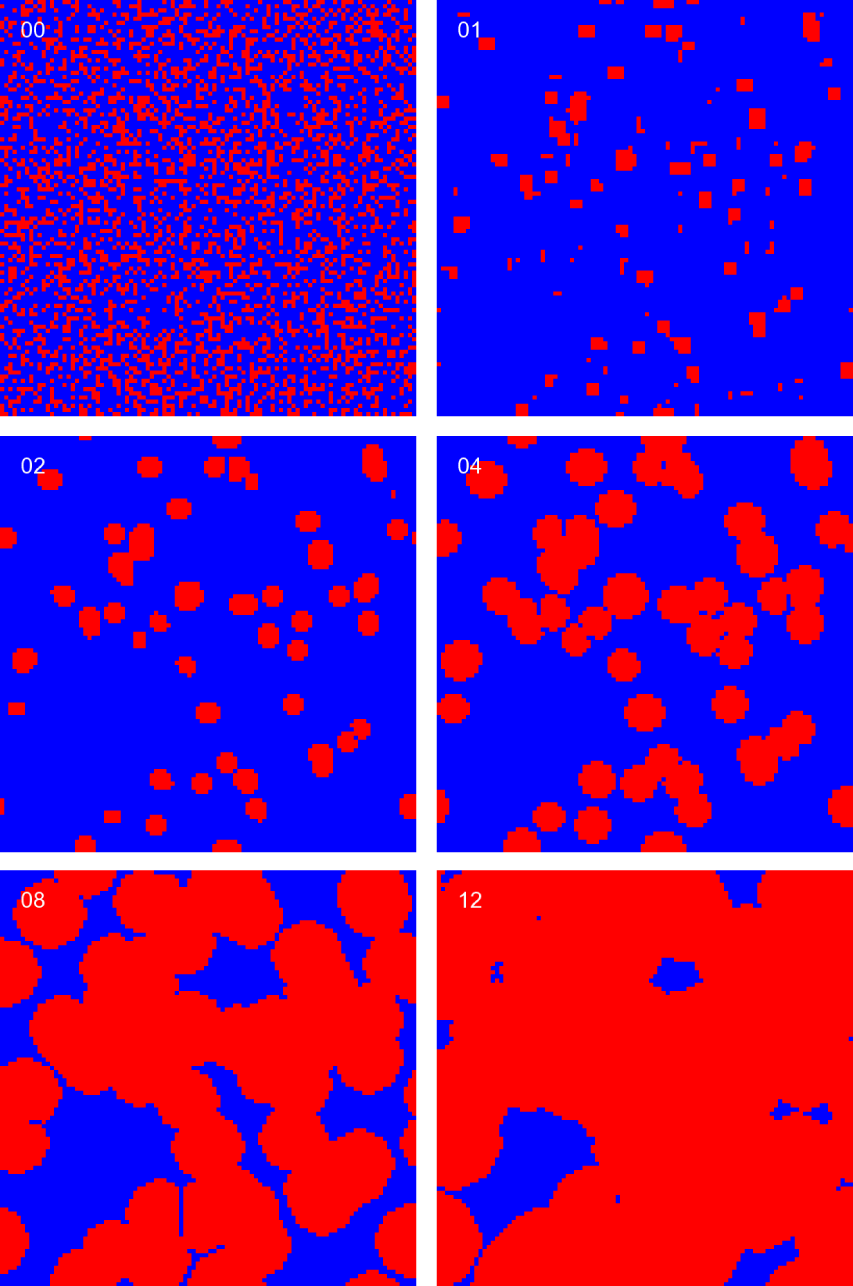}
\caption{(Color online) Snapshots of the evolution of a population on a regular lattice of
degree $k=8$, playing a Stag Hunt game ($S=-0.65$ and $T=0.65$). Cooperators and defectors are displayed as in Fig.~\ref{fig:evol-repl}. The update rule is unconditional
imitation and the initial density of cooperators is $x^0 =1/3$ (this lower value
than that of Fig.~\ref{fig:evol-repl} has been used to make the
evolution longer and thus more easily observable). The upper left label shows
the time step $t$. As with the replicator rule, during the initial time steps clusters emerge from cooperators with high local density of cooperators $\hat{x}$. In
this case, however, the interfaces advance deterministically at each time step, thus producing a much more rapid evolution (compare the
time labels with those of Fig.~\ref{fig:evol-repl}).}
\label{fig:evol-uncimit}
\end{figure*}

\begin{figure*}
\centering
\includegraphics[width=0.99\textwidth]{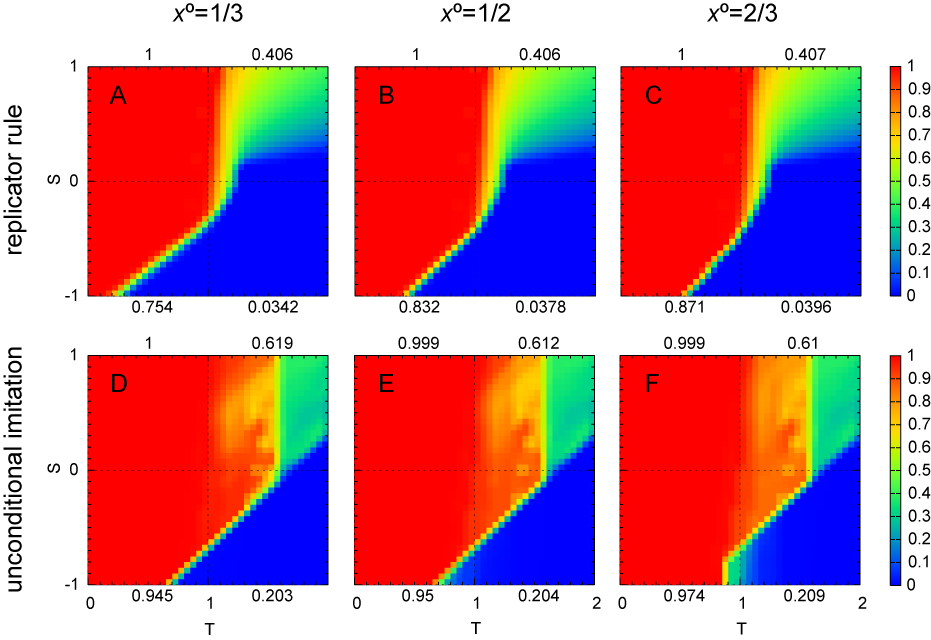}
\caption{(Color online) Asymptotic density of cooperators $x^*$ in a regular lattice of degree
$k=8$, for different initial densities of cooperators $x^0=$ 1/3 (A, D), 1/2 (B,
E) and 2/3 (C, F). The update rules are the replicator rule (upper row, A to C)
and unconditional imitation (lower row, D to F). With the replicator rule, the
evolutionary outcome in Stag Hunt depends on the initial condition, as is
revealed by
the displacement of the transition line between full cooperation and full
defection. However, with unconditional imitation this transition line remains in
the same position, thus showing the insensitivity to the initial condition. In
this case, the outcome is determined by the presence of small clusters of
cooperators in the starting random population, which is ensured for a large
range of values of the initial densities of cooperators $x^0$.}
\label{fig:initial-condition}
\end{figure*}

In other words, for Stag Hunt games the large clustering of the network allows
the high values of local densities caused by random fluctuations in the
initial condition to propagate all over the population. This is the
reason why, in the range of parameters where the population structure
is critical for the prevailing of cooperation, mesoscopic structures
develop in the form of compact clusters of cooperators, as was observed by
Nowak and May \cite{nowak:1992}. It is at the interfaces of these clusters that the explained mechanism takes place. See Fig.~\ref{fig:evol-repl} for some snapshots of a typical example of this phenomenon. It is clear that for these interfaces to propagate all over the population, the nodes that provide the clustering to the network, i.e.\ those that have a number of neighbors that are neighbors themselves, must be connected throughout the full network structure, as it is the case with the spatial networks studied here. Interestingly, this observation is in agreement with the importance of the ``percolation of overlapping triangles'' reported in the context of other network models \cite{vukov:2006}. Considering small-world networks, clusters of cooperators are able to grow like in regular lattices, because of the network clustering. But in this case, the existence of ``shortcuts'' in the network, due to the low mean distance between nodes, facilitates the spreading of clusters, thus reducing the time needed for the population to reach full cooperation.

In the case of Snowdrift games, cooperators tend to aggregate as well, but this immediately raises the payoff of the surrounding defectors more than
that of the cooperators, which makes them switch to defection, thus
disintegrating the embryonic cluster. The overall effect is a
decrease in the global cooperator density. Besides, as the nucleation
effect does not develop beyond its initial stages, the inhibition of
cooperation in Snowdrift games is not as strong as the promotion in Stag Hunt
games.

Nevertheless, unconditional imitation does promote cooperation in Snowdrift
and even in Prisoner's Dilemma, on lattices with large clustering.
Obviously, the effect of the network topology is basically the same as with the stochastic update rules. This sharp difference in results lies in
the lack of stochasticity of unconditional imitation, which makes the
cluster interfaces advance uniformly, as Fig.~\ref{fig:evol-uncimit} illustrates. As a consequence, the dynamics of flat interfaces takes on a special relevance in this case, determining the evolutionary outcome. For example, computing the payoff balance between cooperators and defectors arranged on both sides on a flat interface yields the most important transition line between full cooperation and full defection in the $ST$-plane
(see Figs.~\ref{fig:lattice-uncimit}~E-F): $T-S=2$
for $k=6$, and $T-S=5/3$ for $k=8$, as was reported in \cite{nowak:1992} (see
also \cite{schweitzer:2002}).

An interesting ``rule of thumb'' to estimate the fate of cooperation on a
spatially structured population was proposed by Hauert \cite{hauert:2002,hauert:2001}: cooperation emergence is directly related to the stability and growth of $3 \times 3$ clusters (see also \cite{brauchli:1999} for a discussion on this point). We confirm this rule for unconditional imitation and $k=8$, because in that case the growing conditions of a $3 \times 3$ cluster are exactly the same as the advance conditions of a flat interface mentioned above. Interestingly, this rule implies an independence of the evolutionary outcome from the initial density of cooperators $x^0$. As long as there were, in the initial population, a small cluster that would grow (in fact, with $k=8$, a $2 \times 3$ cluster is sufficient), the population would reach full cooperation. As expected, Fig.~\ref{fig:initial-condition}  shows that this is the case with unconditional imitation (lower row). But with the replicator rule (upper row) the corresponding transition between full cooperation and full defection does depend on the initial density of cooperators. This dependence on the initial condition means that, in the case of stochastic update rules, there is not one or a small subset of privileged configurations that determine the
evolutionary outcome. This fact suggests that techniques such as pair
approximation methods \cite{van-baalen:2000} are more appropriate to obtain
estimations in this case. Notice also that, for stochastic rules and Stag Hunt games, in contrast to the ``leveling out'' of the initial density of cooperators reported by \cite{hauert:2002}, we have found a promotion
of cooperation for all the initial conditions studied, as the first row of
Fig.~\ref{fig:initial-condition} displays in comparison with Fig.~\ref{fig:compnet}.

Finally, the comparison of results between strong and weak selection pressure reveals important qualitative differences: with weak selection the effect of spatial structure is clearly attenuated and the asymmetry between the influence on Stag Hunt and Snowdrift games becomes a symmetric effect. We have seen that the effects with strong selection are rooted in the strong
correlations that appear in the population as clusters form and grow. With
weak selection correlations develop in a completely different manner, hence the change in results. The weak selection limit in the Fermi rule, given by $\beta
\to 0$ in Eq.~(\ref{eq:fermirule}), yields a probability of copying the
strategy of a neighbor very close to 0.5, with a little bias
proportional to the difference of payoffs. As a consequence, the changes
in the local densities that the game causes diffuse over the population,
without directly affecting the fate of the neighborhood that originated
them, and thus giving rise to much weaker correlations than in the strong
selection case.

\section{Conclusions}

\label{sec:conclusions}

In this work we have addressed some important open questions about the effect
of spatial structure on the evolution of cooperation. We have found an
unquestionable dependence of the evolutionary outcome on the update rule, which
has in turn consequences on the robustness of the spatial effects and on the
influence of the synchronicity of the updating. Coordination or Stag Hunt games
have showed up as the prototypical games for the positive effect of spatial
structure on the evolution of cooperation. The importance of network clustering
as a general property has been clarified, along with its role in the influence
of small-world networks. Selection pressure has also been identified as a key
factor in these models, with a clear qualitative and quantitative influence,
related to the symmetry of effects on coordination and anti-coordination
games.

Methodologically, our work makes it clear the interest of studying the
two-dimensional $ST$-space of $2 \times 2$ games, beyond a particular case or a
one-dimensional parametrization of a game. Recent work on different topics has
also shown so, as for example that of Santos and co-workers on the issue of
network heterogeneity and the scale-free property
\cite{santos:2006a,santos:2006b}.

To conclude, we must recognize the strong dependence on details of evolutionary games on spatial networks. As a consequence, it does not seem plausible to expect general laws that could be applied in a wide range of practical settings. On the contrary, a close modeling including the kind of game, the evolutionary dynamics and the population structure of the concrete problem seems mandatory to reach sound and compelling conclusions. With no doubt this is an enormous challenge, but we believe that this is one of the most promising paths that the community working in the field can explore.

\appendix*

\section{Methods information}

\label{ap:methods}

All the simulations were performed for a population size of $N = 10^4$. The
initial density of cooperators was $x^0 = 0.5$ and the update of strategies was done synchronously, unless otherwise stated.

With synchronous update, all the individuals in the
population play the game once with all their neighbors, compare payoff with
them and decide the new strategy for the next time step. Then, they all update
their strategy at once and their payoff is set to zero before the next step.
With the asynchronous update, an individual is chosen at random. She and her
neighbors play the game once, each one with all her neighbors, so that they
earn the same payoff that they would have earned with a synchronous update.
Then, the chosen individual compares payoff with her neighbors and updates her
strategy accordingly. Finally, the payoff of all the individuals is set to zero before the next time step.

The time of convergence in the simulations was $T = 10^4$ steps for
synchronous update and $T=N \times 10^4$ for the asynchronous case ($N$ is the
population size). This way the total number of update events is the same for
both schemes. If the population did not reach full cooperation or defection, an average of the cooperator density during the last tenth of the time evolution
was used to obtain the asymptotic cooperator density. Figure~\ref{fig:async-2}
shows that this time of convergence is enough to reach a steady state, specially
in the case of stochastic update rules, which are much slower than the
deterministic unconditional imitation rule. Notice that we have used a much
larger time of convergence than \cite{hauert:2002}, which employed 48 time steps
for a population size of $51 \times 51$. We have found that, for population
sizes like these ones, times well over $10^3$ steps are needed to ensure a
correct convergence, in agreement with many other works in the field, like for
example
	\cite{brauchli:1999,cohen:2001,tang:2006,perc:2008,tomassini:2006,vukov:2006}. Hauert justified in \cite{hauert:2002} the choice of the time of convergence to minimize finite-size effects, so the system did not become aware of its finite dimensions. We disagree with this argument. Long times of convergence are typically needed for Stag Hunt games near the transition between full cooperation and full defection (see Fig.~\ref{fig:async-2}). In these cases, clusters of cooperators may require a long time to grow and occupy all the population. This means that in order to reach the steady state interactions of the system with its periodic images are unavoidable. On the other hand, the influence of system size manifests itself as a lower or greater probability of favorable configurations at initial time, which can logically  have an effect on the final outcome (see Sec.~\ref{sec:discussion}). So to actually determine finite-size effects there is no other way to proceed but to increase the size of the system. Reducing the evolution time only introduces uncontrolled errors of lack of convergence.

The studied region in the $ST$-plane was sampled in steps of 0.05. For each
point in the resulting $41 \times 41$ grid, which corresponds to a concrete
game, 100 realizations were performed to obtain a final average value for the
asymptotic density of cooperators. The mean cooperation index for each game was
calculated from the asymptotic values with the standard two-dimensional
Simpson's quadrature rule \cite{stoer:2002}.

Each realization started from a newly generated population, with strategies
randomly assigned and the network, when applicable, also randomly built. The
homogeneous random networks were constructed directly, assigning links
randomly in the population, while ensuring an equal number of links for every
individual. All the regular lattices were built with periodic boundary
conditions. Regular lattices of degrees $k=4$ and $k=8$ were built with two-dimensional square grids and, respectively, von Neumann and Moore neighborhoods. Regular lattices of degree $k=6$ were built with two-dimensional triangular grids and the 6 nearest neighbors.

\begin{acknowledgments}
We thank Alex Arenas for a critical reading of the manuscript. This work is
partially supported by Ministerio de Educaci\'on y Ciencia (Spain) under Grants Ingenio-MATHEMATICA and MOSAICO, and by Comunidad de Madrid (Spain) under Grants SIMUMAT-CM and MOSSNOHO-CM.
\end{acknowledgments}

\bibliography{evol-coop}

\end{document}